\documentclass[]{spie}
\usepackage{graphicx,color}
\usepackage{lscape}
\usepackage{bm}

\title{Statistics of Current Fluctuations and Electron-Electron Interactions in
Mesoscopic Coherent Conductors}

\author{Dmitri S. Golubev\supit{a,c}, Artem V. Galaktionov\supit{b,c},
and Andrei D. Zaikin\supit{b,c}
\skiplinehalf
\supit{a}Institut f\"ur Theoretische Festk\"orperphysik,
Universit\"at Karlsruhe, 76128 Karlsruhe, Germany; \\
\supit{b}Forschungszentrum Karlsruhe, Institut f\"ur Nanotechnologie,
76021, Karlsruhe, Germany; \\
\supit{c}P.N. Lebedev Physics Institute, 119991 Moscow, Russia
}

\begin{document}
\maketitle
\begin{abstract}
We formulate a general path integral approach which describes statistics of
current fluctuations in mesoscopic coherent
conductors at arbitrary frequencies and in the presence of
interactions. Applying this approach to the non-interacting case, we
analyze the frequency dispersion of the third cumulant of the
current operator ${\cal S}_3$ at frequencies well below both
the inverse charge relaxation time and the inverse electron dwell
time. This dispersion turns out to be important in the frequency
range comparable to applied voltages. For comparatively transparent conductors
it may lead to the sign change
of ${\cal S}_3$.
We also analyze the behavior of the second cumulant of the current operator 
${\cal S}_2$ (current noise) in the presence of electron-electron interactions.
In a wide range of parameters we obtain explicit
universal dependencies of ${\cal S}_2$ on temperature,
voltage and frequency.
We demonstrate that Coulomb interaction decreases the
Nyquist noise. In this case the interaction correction
to the noise spectrum is governed by the combination
$\sum_nT_n(T_n-1)$, where $T_n$ is the transmission of the
$n$-th conducting mode. The effect
of electron-electron interactions on the shot noise is more
complicated. At sufficiently large voltages we recover two
different interaction corrections entering with opposite signs.
The net result is proportional to $\sum_nT_n(T_n-1)(1-2T_n)$,
i.e. Coulomb interaction decreases the shot noise at low transmissions
and increases it at high transmissions.
\end{abstract}

\section{Introduction}

Experimental and theoretical
studies of mesoscopic conductors reveal a rich variety of low temperature
properties and effects caused by interplay between scattering,
interactions, quantum coherence and electron charge discreteness.
Investigations of the whole scope of these effects
are of primary interest because of their fundamental importance
as well due to rapidly growing number of their potential applications.

A great deal of information is usually obtained from studying electron
transport. Additional information can be extracted
from investigations of fluctuation effects. During last years much attention
has been devoted to the shot noise \cite{bb} described by the second
moment of the current operator. One can also study higher order
correlators of the current operator thereby
extending the amount of information already obtained from electron transport
and shot noise. Recently the first experimental study
of the third current cumulant in mesoscopic tunnel
junctions was reported \cite{reul}.

A theoretical framework which enables one to analyze
statistics of charge transfer in mesoscopic conductors
was developed in Ref. 3.
This theory of full counting statistics (FCS) allows to evaluate any
cumulant of the current operator in the absence of interactions and
in the zero frequency limit. Under these conditions higher order current
cumulants were investigated by a number of authors
\cite{LLY,lev2,Nag,GG}. In order to include interactions and to
analyze frequency dispersion of current fluctuations it is
necessary to go beyond the FCS theory and to develop a more general
real time path integral technique \cite{GZ00,GGZ,GGZ2,KN,BN}.

The goal of the present paper is to address statistics of current
fluctuations at non-zero frequencies and in the presence of electron-electron
interactions. In what follows we will develop a general path integral
approach which allows to perform a complete analysis of electron-electron
interaction effects in mesoscopic coherent conductors
described by an arbitrary -- though energy independent -- scattering matrix.
This approach also provides a straightforward generalization of the FCS theory
 \cite{lev1} to non-zero frequencies. Of central importance in this context
is a general and formally exact expression for the real time
effective action of a coherent conductor. This expression will be
derived and analyzed in various limiting cases in Section II
of the paper. With the aid of our technique in Section III we
will investigate the frequency dispersion of the third cumulant of
the current operator in mesoscopic coherent conductors. Section IV is devoted
to the analysis of the electron-electron interaction correction to the
second cumulant of the current operator, i.e. to the current noise.
The paper is briefly concluded in Section V.

\section{The model and effective action}

\subsection{General analysis}

In our analysis we will use the real time path integral formalism
developed for the systems of interacting fermions \cite{GZ97}. After
the standard Hubbard-Stratonovich decoupling of the interaction term
in the Hamiltonian one can exactly integrate out fermions and arrive
at the effective action $S$ which depends on the fluctuating fields
$V_{1,2}(t,\bm{r})$. Let us define
\begin{equation}
{\rm e}^{iS_0}=
{\rm Tr}\left[{\cal T}{\rm e}^{-i\int_0^t dt'{\bm H}_1(t')}
\hat{\bm \rho}_0 \tilde {\cal T}{\rm e}^{i\int_0^t dt'{\bm H}_2(t')}\right],
\label{eiS}
\end{equation}
with the trace taken over the fermionic variables. Here $\hat{\bm \rho}_0$ is
the initial $N-$particle density matrix of electrons,
\begin{eqnarray}
{\bm H}_{1,2}=\sum_{\sigma}\int d^3{\bm
r}\,\hat\Psi^\dagger_\sigma({\bm r}) \hat
H_{1,2}(t)\hat\Psi_\sigma({\bm r}), \;\;\;\;\;\; \hat
H_{1,2}(t)=-\frac{\nabla^2}{2m}+U(\bm{r})-eV_{1,2}(t,\bm{r}).
\label{H}
\end{eqnarray}
Here ${\bm H}_{1,2}$ are the effective Hamiltonians on the forward
and backward parts of the Keldysh contour, $U(\bm{r})$ describes
the static potential, ${\cal T}$ and $\tilde{\cal T}$ are
respectively the forward and backward time ordering operators.
Integrating out fermions in Eq. (\ref{eiS}), we obtain
\begin{equation}
iS_0=2\,{\rm Tr}\ln[1+(\hat u_2^{-1}\hat u_1-1)\hat\rho_0],
\label{S}
\end{equation}
where \begin{equation} \hat u_{1,2}(t)={\cal
T}\exp\left(-i\int_0^t dt' \hat H_{1,2}(t')\right)
\end{equation}
are the evolution operators pertaining to the Hamiltonians
(\ref{H}), $\hat\rho_0$ stands for the initial single-particle
density matrix.

The total action of the problem $S$ also includes the contribution
of the fluctuating electromagnetic field outside the scatterer.
This contribution reads
\begin{eqnarray}
iS_{\rm ex}&=&-\frac{i}{e^2}\int_0^t dt_1dt_2\,\varphi^-_S(t_1)\bigg(\int\frac{d\omega}{2\pi} \frac{{\rm e}^{-i\omega (t_1-t_2)}}{Z_S(\omega)}\bigg)
\dot\varphi^+_S(t_2)
\nonumber\\ &&
-\,\frac{1}{2 e^2}\int_0^t dt_1dt_2\,\varphi^-_S(t_1)\bigg(\int\frac{d\omega}{2\pi} {\rm Re}\bigg[\frac{1}{Z_S(\omega)}\bigg]
\omega\coth\frac{\omega}{2T}{\rm e}^{-i\omega (t_1-t_2)}\bigg)
\varphi^-_S(t_2).
\label{Sex}
\end{eqnarray}
Here $Z_S(\omega)$ is the effective linear impedance of the
electromagnetic environment, and $\varphi_S$ is the quantum phase variable 
related to the fluctuating voltage $V_S$ across the
the external leads by means of the equation  $\dot \varphi_S=eV_S$. 
The superscripts ($\pm$) just reflect the fact that the phase variable is defined
on the Keldysh contour, so that symmetric ($\varphi_S^+$) and antisymmetric
($\varphi_S^-$) combinations of the phases on two branches of this contour
should be introduced.

Although the above expression is valid for
arbitrary  $Z_S(\omega)$, for the sake of definiteness and
simplicity below we will choose $1/Z_S(\omega)=-i\omega C+1/R_S,$
where $C$ is the barrier capacitance  and $R_S$ is the resistance
of external leads (see Fig. 1). The total action of our system
then reads
$$
S=S_{\rm ex}+S_0.
$$
\begin{figure}
\begin{center}
\includegraphics[width=7cm]{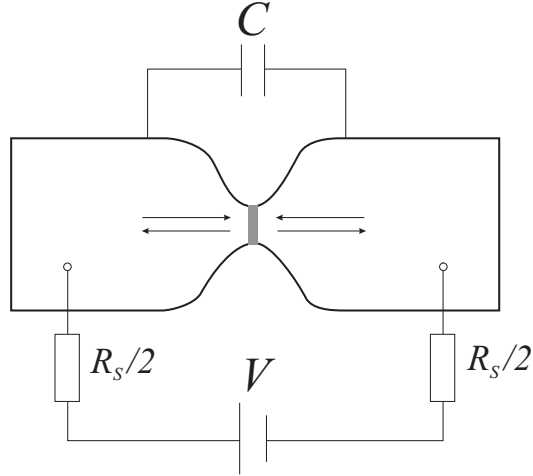}
\end{center}
\caption{Schematics of the system under consideration. The scatterer
is placed in-between two big metallic reservoirs and is characterized
by the Landauer conductance $1/R$ and the effective capacitance $C$. The
reservoirs are connected to the voltage source via leads with the total
Ohmic resistance $R_S$.}
\end{figure}
In order to evaluate the evolution operators $\hat u_{1,2}$ and
derive the expression for the action (\ref{S}) it is necessary to
specify the model of a mesoscopic conductor. Here we will adopt
the standard model of a (comparatively short) coherent conductor
placed in-between two bulk metallic reservoirs, see Fig. 1. The
electron dwell time $\tau_D$ is supposed to be shorter than any
relevant time scale in our problem. Energy and phase relaxation
times are, on the contrary, assumed to be long, i.e. inelastic
relaxation is allowed in the reservoirs but not inside the
conductor. Under these assumptions electron transport through the
conductor can be described by the energy independent scattering
matrix
\begin{equation}
\hat S = \left( \begin{array}{cc} \hat r & \hat t'
\\ \hat t & \hat r'
\end{array}\right)
\label{sm}
\end{equation}
and the standard Landauer formula for the conductance $1/R=(e^2/\pi)\sum_nT_n=
(e^2/\pi){\rm Tr} \hat t^\dagger\hat t$ can be used, where $T_n$ defines
the transmission for the $n$-th conducting mode.

The effective action (\ref{S})
can be expressed via the fluctuating phase fields $\varphi_{1,2}$
which are in turn related to the jumps of the fields $V_{1,2}$ across
the scatterer as $\dot \varphi_{1,2}=e(V_{L1,2}-V_{R1,2})$, where
$V_{L,R}$ are fluctuating in time but constant in space fields in the left and right
reservoirs. We note that in this case the right hand side
of Eq. (\ref{eiS}) differs from the FCS generating functional \cite{KN2} only by a gauge transformation.

Within the above model the evolution operators $\hat u_{1,2}$ were
derived in Refs. 8 and 9. Combining these expressions
with Eq. (\ref{S}) after some algebra we find \cite{GGZ2}
\begin{eqnarray}
iS_0=2{\rm Tr}\ln\left\{\hat 1 +\theta(t-x)\theta(x)
\left[\begin{array}{cc} \hat t^\dagger\hat t({\rm e}^{i\varphi^-(x)}-1)
& 2i \hat t^\dagger\hat
r' \sin\frac{\varphi^-(x)}{2} \\  2i \hat r^{\prime \dagger}\hat t
\sin\frac{\varphi^-(x)}{2} & \hat t^{\prime \dagger}\hat t'( {\rm e}^{
-i\varphi^-(x)}-1)
\end{array}
\right]
\left[\begin{array}{c}
\rho_0(y-x){\rm e}^{i\frac{\varphi^+(x)-\varphi^+(y)}{2}}\hspace{0.9cm}  0
\\  0\hspace{0.9cm}  \rho_0(y-x){\rm e}^{i\frac{\varphi^+(y)-\varphi^+(x)}{2}}
\end{array}
\right] \right\}.
\label{S2}
\end{eqnarray}
Here we introduced $\varphi^+= (\varphi_{1} +\varphi_{2})/2$,
$\varphi^-= \varphi_{1} -\varphi_{2}$ and
\begin{equation}
\rho_0(x)=\int \frac{dE}{2\pi}\frac{e^{iEx}}{1+e^{E/T}}=\frac{1}{2}\delta(x)-\frac{iT}{2\sinh\pi Tx}.
\end{equation}
Taking the trace in Eq. (\ref{S2}) implies convolution with respect to
internal time variables ($x$ and $y$). The total time span is
denoted by $t$. 

Eq. (\ref{S2}) defines a formally exact effective action for a
coherent conductor described by an arbitrary energy independent
scattering matrix (\ref{sm}). This expression allows to fully
determine statistics of current fluctuations at arbitrary
frequencies and in the presence of interactions. In the case of
equilibrium fluctuations the formula (\ref{S2}) represents a real
time analogue of the effective action \cite{SZ89,SZ90,Naz} derived
within the Matsubara technique. A formula similar to Eq.
(\ref{S2}) was also presented recently in Refs. 11 and 12. In
addition we note that, provided the field $\varphi^-$ does not
depend on time, Eq. (\ref{S2}) coincides with the generating
function considered, e.g., in the problem of adiabatic pumping
through mesoscopic conductors \cite{AKL}.

In order to proceed with further calculations it is useful to
investigate the properties of the general expression (\ref{S2})
and -- whenever it is possible -- to find its simplified
representations. This goal can be achieved with the aid of several
different approximations which can be appropriate in the
corresponding physical limits. Some of these approximations and
the resulting expressions for the effective action are specified
below.

\subsection{Expansion in $T_n$}

Provided channel transmissions $T_n$ of the scatterer are
sufficiently low it can be useful to employ a regular expansion of
the action (\ref{S2}) in powers of $T_n$. The first order terms of
this expansion just yield the standard AES action for tunnel
barriers \cite{SZ90}. Here we will proceed further with this
expansion and establish the second order in $T_n$ contributions to
$S_0$. In order to recover all such terms it is necessary to
expand the logarithm in Eq. (\ref{S2}) up to the fourth order.
After a straightforward -- although somewhat tedious -- calculation we
obtain
\begin{eqnarray}
 iS_0&=&-\frac{ig}{2\pi}\int_0^{t} dx\;
\dot\varphi^+(x)\sin\varphi^-(x)\left[1+\frac{2(1-\beta)}{3}\sin^2\frac{\varphi^-(x)}{2}\right]
\nonumber\\ &&
-4g\int_{0}^{t}dx_1
\int_{0}^{t}dx_2\,\rho_0^2(x_1-x_2)\sin\frac{\varphi^-(x_1)}{2}
\sin\frac{\varphi^-(x_2)}{2}
\nonumber\\ &&\times\,
\left\{ \beta\cos[\varphi^+(x_1)-\varphi^+(x_2)]+
(1-\beta)\cos\frac{\varphi^-(x_1)-\varphi^-(x_2)}{2}
 \right\}
\nonumber\\ &&
+\frac{2i g T^3(1-\beta)}{3}\int_0^{t} dx_1 \int_0^{t} dx_2\int_0^{t}
dx_3\,\frac{\sin\frac{\varphi^-(x_1)}{2}
\sin\frac{\varphi^-(x_2)}{2} \sin\frac{\varphi^-(x_3)}{2}}{\sinh\pi T(x_2-x_1)\sinh\pi T(x_3-x_2)\sinh\pi T(x_1-x_3)}
\nonumber\\ &&
\times \left\{ \sin[\varphi^+(x_2)-\varphi^+(x_1)]\cos\frac{\varphi^-(x_3)}{2}+
\sin[\varphi^+(x_3)-\varphi^+(x_2)]\cos\frac{\varphi^-(x_1)}{2}
\right.
\nonumber\\ &&
\left.
+\,\sin[\varphi^+(x_1)-\varphi^+(x_3))\cos\frac{\varphi^-(x_2)}{2}\right\}
\nonumber\\ &&
-8g(1-\beta) \int_0^{t} dx_1 \int_0^{t} dx_2
\int_0^{t} dx_3 \int_0^{t} dx_4\, \rho_0(x_2-x_1)
\rho_0(x_2-x_3)\rho_0(x_4-x_3)\rho_0(x_4-x_1)
\nonumber\\ &&  \times
\sin\frac{\varphi^-(x_1)}{2} \sin\frac{\varphi^-(x_2)}{2}
\sin\frac{\varphi^-(x_3)}{2}\sin\frac{\varphi^-(x_2)}{2}\cos\left[
\varphi^+(x_1)-\varphi^+(x_2)+\varphi^+(x_3)-\varphi^+(x_4) \right].
\label{aes2}
\end{eqnarray}
Here we introduce the dimensionless conductance of the scatterer
$g=2\sum_n T_n$ as well as its Fano factor $\beta=\sum_nT_n(1-T_n)
/\sum_n T_n$.  Eq. (\ref{aes2}) represents a complete expression
for the effective action valid up to the second order in the
transmissions $T_n$. This expression involves no further
approximations and fully accounts for the non-linear dependence on
the fluctuating phase fields $\varphi^\pm$.

\subsection{Reflectionless barriers}

Another physically important limit is that of reflectionless barriers 
$\hat r=0$. In this limit the general expression (\ref{S2}) can be 
reduced to a very simple form by means of the exact procedure which we outline
below. The derivation is based on the following property of the Fermi function
$n(E)=1/(1+\exp(E/T)):$
\begin{equation}
n(E)n(E+\omega)=\frac{1}{2}\left(1-\coth\frac{\omega}{2T}\right)n(E)
+\frac{1}{2}\left(1+\coth\frac{\omega}{2T}\right)n(E+\omega).
\label{id1}
\end{equation}
This relationship plays the same role as the Ward identity in the Luttinger 
liquid theory. Rewriting (\ref{id1}) in the coordinate representation one finds
\begin{equation}
\rho_0(x-y)\rho_0(y-z)=(f(x-y)+f(y-z))\rho_0(x-z),
\label{id}
\end{equation}
where
$
f(x)=\int\frac{d\omega}{2\pi}\,\frac{1}{2}\left(\coth\frac{\omega}{2T}+1\right)\,
{\rm e}^{-i\omega x}=-\frac{iT}{2}{\cal P}\coth[\pi Tx]+
\frac{1}{2}\delta(x).
$
Let us formally expand the action (\ref{S2}) in powers of the transmission
matrix $\hat t^\dagger\hat t$. This expansion gives rise to the integrals of the type
\begin{equation}
J_N=\int_0^t dx_1dx_2\dots dx_N\; a(x_1)\rho_0(x_1-x_2)a(x_2)\rho_0(x_2-x_3)a(x_3)\dots \rho_0(x_{N-1}-x_N)a(x_N)\rho_0(x_N-x_1),
\label{JN}
\end{equation}
where $a(x)={\rm e}^{-i\varphi^-(x)}-1$.
The identity (\ref{id}) permits to evaluate such integrals exactly.
It will be convenient for us to carry out $N-1$ permutations of 
the variables $x_j$ as follows.
We start from the set (1) $x_1,x_2,x_3\dots x_N,$ and further introduce
the following sets of integration parameters
\begin{eqnarray}
\begin{array}{ccccccc}
(2) & x_1 & x_N & x_2 & x_3 & \dots & x_{N-1} \\
(3) & x_1 & x_2 & x_N & x_3 & \dots & x_{N-1} \\
\dots &&&&&& \\
(N-1) & x_1 & x_2 & \dots  & x_{N-2}& x_N & x_{N-1},
\end{array}
\end{eqnarray}
i.e.. we successively insert $x_N$ in-between $x_1$ and $x_2$, $x_2$
and $x_3$, an so on.  Symmetrizing the integral (\ref{JN}) with respect to
these $N-1$ permutations and excluding $x_N$ from the functions $\rho_0$
by virtue of the identity (\ref{id}), we get
\begin{eqnarray}
J_N=\frac{1}{N-1}\int dx_1dx_2\dots dx_{N-1}\;\; a(x_1)a(x_2)\dots a(x_{N-1})
\big(a(x_1)+a(x_2)+\dots +a(x_{N-1})\big)
\hspace{2.7cm}
\nonumber\\ \times\,
\rho_0(x_1-x_2)\rho_0(x_2-x_3)\dots \rho_0(x_{N-1}-x_1)
+\frac{1}{2\pi(N-1)}\int\frac{dx_1dx_2}{2\pi}\,\frac{\pi^2T^2}{\sinh^2\pi T(x_1-x_2)}
a^{N-1}(x_1)a(x_2).
\label{JN-1}
\end{eqnarray}
Eq. (\ref{JN-1}) allows one to relate $J_N$ and $J_{N-1}$. 
After $N-2$ iterations one finds   
the following result:
\begin{eqnarray}
J_N=\int_0^t dx\,\rho_0(x,x)a^N(x)-\frac{1}{4}\int_0^t\frac{dx dy}{2\pi}\,\left(-\frac{1}{\pi}\frac{\pi^2T^2}{\sinh^2\pi T(x-y)}\right)
\sum_{k=1}^{N-1}\frac{N}{k(N-k)}a^k(x)a^{N-k}(y).
\label{JNN}
\end{eqnarray}
Now we are in a position to evaluate the trace
$A={\rm Tr}\ln\big[\hat 1+\theta(t-x)\theta(x)
  a(x)\rho_0(x-y)\big]$. Exploiting Eq. (\ref{JNN}) we obtain
\begin{eqnarray}
A=\int_0^t dx\,\rho_0(x,x)\ln(1+a(x)) 
+\frac{1}{8\pi}\int_0^t{dx dy}\,\left(-\frac{1}{\pi}\frac{\pi^2T^2}{\sinh^2\pi T(x-y)}\right)
\ln(1+a(x))\ln(1+a(y)).
\label{A}
\end{eqnarray}
Now we recall that in the case of reflectionless barriers ($T_n=1$ for all
$n$) the action (\ref{S2}) reads
\begin{eqnarray}
iS_0=2N_{\rm ch}\big\{ {\rm Tr}\ln\big[\hat 1+\theta(t-x)\theta(x) a(x)\rho_R(y,x)\big]
+{\rm Tr}\ln\big[\hat 1+\theta(t-x)\theta(x) a^*(x)\rho_L(y,x)\big]\big\},
\end{eqnarray}
where $N_{\rm ch}$ is the total number of fully open channels in the conductor
and $\rho_{L,R}(y,x)={\rm e}^{i[\varphi^+_{L,R}(x)-\varphi^+_{L,R}(y)]}\rho_0(y-x)$. 
Expanding this action in powers of $a(x)$ and $a^*(x)$ and applying the formula
(\ref{A}) we observe
that the factors ${\rm e}^{\varphi^+_{L,R}(x)-\varphi^+_{L,R}(y)}$ 
cancel out exactly in all orders except in the first term of Eq. (\ref{A}).
This remaining term should be treated with sufficient 
care having in mind the formal divergence $\rho_0(x,y)\to\infty$ at $x\to y$. 
Finally we obtain
\begin{eqnarray}
iS_0=-\frac{ig}{2\pi}\int_0^t dx\,\dot\varphi^+(x)W\big(\varphi^-(x)\big)
+\frac{g}{4}\int_0^t{dx dy}\,\frac{T^2}{\sinh^2\pi T(x-y)}
W\big(\varphi^-(x)\big)W\big(\varphi^-(y)\big),
\label{Sopen}
\end{eqnarray}
where $g=2N_{\rm ch}$ and $W(\varphi^-)$ is the $2\pi-$periodic function equal
to $W(\varphi^-)=\varphi^-$ for $-\pi<\varphi^-<\pi$. Note that for the values
of $\varphi^-$ within the
latter interval Eq. (\ref{Sopen}) exactly coincides with the action for an
Ohmic conductor. Finally we point out that the existence of the exact solution
(\ref{Sopen}) in the limit $\hat r=0$ is not at all surprising since in that
limit the problem can also be exactly treated by means of the bosonization
technique \cite{Matveev}.


\subsection{Local in time part of the action}

The above results for $S_0$ demonstrate that the effective action
can be split into two parts,
\begin{equation}
S_0=S_{\rm local}+S_{\rm nonlocal},
\label{l-nl}
\end{equation}
determined respectively by local and non-local in time
expressions. One can also observe that the term $S_{\rm local}$ is
linear in $\dot\varphi^+$. Such splitting can be performed 
for arbitrary transmission values. It turns out that the local
part of the effective action can be found exactly in all orders in
$T_n$.

In order to derive this exact expression let us first fix
$\varphi^+=eVt$ and $\varphi^-=const$. It is easy to demonstrate
\cite{GGZ2} that under these restrictions the action $S_0$ reduces
to the FCS generating function \cite{lev1}
\begin{equation}
iS_0 \to -\frac{Tt}{\pi}\sum_n\left[v^2-{\rm
Arccosh}^2\left(T_n\cosh(v-i\varphi^-)+(1-T_n)\cosh
v\right)\right], \label{FCS}
\end{equation}
where ${\rm Arccosh}\, x=\ln(x+\sqrt{x^2-1})$ and $v=eV/2T$. The
part $S_{\rm local}$ can be identified from this generating
function in the limit $v\ll 1$. Keeping only linear in $v$ terms
one can rewrite Eq. (\ref{FCS}) in the form
\begin{equation}
iS_0\to -\frac{ieVt}{\pi}
\sum_n\frac{T_n\sin\varphi^-\left(\frac{\pi}{2}-\arctan\frac{1-2T_n\sin^2\frac{\varphi^-}{2}}
{2\sin\frac{|\varphi^-|}{2}\sqrt{T_n\left(1-T_n\sin^2\frac{\varphi^-}{2}\right)}}\right)}
{2\sin\frac{|\varphi^-|}{2}\sqrt{T_n\left(1-T_n\sin^2\frac{\varphi^-}{2}\right)}}+{\cal
O}(v^2).
\end{equation}
Since $S_{\rm local}$ should be determined by the
local in time combination one can now replace $eV$ by an arbitrary
function of time $\dot\varphi^+$ and also allow for arbitrary
changes of $\varphi^-$ in time. This observation immediately
brings us to the final result
\begin{equation}
iS_{\rm local}=-\frac{i}{\pi}\int_0^t dt'\,\dot\varphi^+
\sum_n\frac{T_n\sin\varphi^-\left(\frac{\pi}{2}-\arctan\frac{1-2T_n\sin^2\frac{\varphi^-}{2}}
{2\sin\frac{|\varphi^-|}{2}\sqrt{T_n\left(1-T_n\sin^2\frac{\varphi^-}{2}\right)}}\right)}
{2\sin\frac{|\varphi^-|}{2}\sqrt{T_n\left(1-T_n\sin^2\frac{\varphi^-}{2}\right)}}.
\label{exloc}
\end{equation}
Unfortunately the above
exact expression for $S_{\rm local}$ alone is not
sufficient for a full description of quantum properties of our
system and no such simple derivation procedure exists for the
non-local in time part $S_{\rm nonlocal}$. Nonetheless, Eq. (\ref{exloc}) 
provides an important piece of information about the properties of the
effective action $S_0$ and it can be successfully employed in further
calculations.

\subsection{Expansion in $\varphi^-$}

In a large number of physical situations it is sufficient to
restrict the analysis to the limit of large scatterer conductances
$g \gg 1$. Either in this limit or, equally, in the case $g_S =2\pi/e^2R_S \gg
1$, fluctuations of the quantum field $\varphi^-$ are considerably
suppressed almost at all energies/frequencies. In this case it is
useful to expand the exact expression for $S_0$ in powers of
$\varphi^-$ keeping the full non-linearity in $\varphi^+$ in all
orders of this expansion. Below we will present the result of this
expansion up to the third order in $\varphi^-$. The corresponding
derivation has been worked out in Ref. 9. It yields
\begin{eqnarray}
iS_0&=&
-\frac{ig}{2\pi}\,\int dx\, \left(\varphi^-(x)-\frac{\beta}{6}[\varphi^-(x)]^3\right)\dot\varphi^+(x)
\nonumber\\ &&
+\,\frac{g}{4\pi^2}\int dx_1
dx_2\,\frac{\pi^2 T^2}{\sinh^2\pi T(x_1-x_2)}
\varphi^-(x_1)\varphi^-(x_2)\left[1-\beta+\beta \cos \left(
\varphi^+(x_1)-\varphi^+(x_2)\right) \right]
\nonumber\\ &&
+\,\frac{\pi i\gamma T^3}{6e^2R}\int
dx_1  dx_2 dx_3\frac{\varphi^-(x_1)
\varphi^-(x_2)\varphi^-(x_3)}{\sinh[\pi T(x_{2}-x_1)]\sinh[\pi T(x_{3}-x_2)] \sinh[\pi
T(x_{1}-x_3)]}
\nonumber\\ &&
\times\, \left\{\sin[\varphi^+(x_2)-\varphi^+(x_1)]+
\sin[\varphi^+(x_3)-\varphi^+(x_2)]
+
\sin[\varphi^+(x_1)-\varphi^+(x_3)]
   \right\},
\nonumber
\label{S3}
\end{eqnarray}
where we have defined $ \gamma={\sum T_n^2(1-T_n)}/{\sum T_n}$.
Obviously, the local in time part of Eq. (\ref{S3}) can also be
recovered by a direct expansion of Eq. (\ref{exloc}) in powers of
$\varphi^-$.

Below we will make use of the result (\ref{S3}) in order to
evaluate the frequency dependence of the third cumulant of the
current operator and to derive the interaction correction to the
shot noise spectrum.

\section{Frequency dependence of the third current cumulant}

To begin with, let us recall that the standard way to describe 
the current noise in our system is
to evaluate the symmetrized correlation function
\begin{equation}
 {\cal S}(t,t')=\frac{1}{2}\langle \hat I(t)\hat I(t')+\hat I(t')\hat I(t)
 \rangle -\langle \hat I \rangle^2,
\label{corr}
\end{equation}
where $\hat I$ is the current operator in the circuit of Fig. 1. After the
Fourier transformation within our path integral technique 
the same quantity can be expressed as follows
\begin{equation}
{\cal S}_2(\omega) =-e^2 \int d\tau\,{\rm e}^{i\omega\tau}
\int {\cal D}\varphi_1{\cal D}\varphi_2 \frac{\delta^2 }{\delta\varphi^-(t+\tau)\delta\varphi^-(t)}
\;{\rm e}^{iS[\varphi^+,\varphi^-]}
\label{noise1}
\end{equation}
Here the path integral is taken over all possible configurations of the phases
$\varphi^\pm$. In this way the effect of Coulomb interaction on current noise
is fully accounted for by Eq. (\ref{noise1}). This effect will be studied in
the next section.

Proceeding along the same lines one can also define higher cumulants
of the current operator. Below we will only analyze the frequency dependence 
of the third cumulant. In doing so, we will ignore interactions. In this
case the path integral is dominated by the trajectories 
$\varphi^+(t)=eVt$ and $\varphi^-=0$, the latter condition being imposed 
after taking the functional derivatives:
\begin{equation}
{\cal S}_3(\omega_1,\omega_2)=e^3\int d\tau_1d\tau_2\,{\rm e}^{i\omega_1 \tau_1} {\rm
e}^{i\omega_2 \tau_2}\;  \left.\frac{\delta^3
S}{\delta\varphi^-(t)\delta\varphi^-(t-\tau_1)\delta\varphi^-(t-\tau_2)}\right|_{\varphi^-=0}.
\label{cum}
\end{equation}
It is important to emphasize that the current through the scatterer should 
be treated as a quantum operator. Since in general the current operators 
do not commute if taken at different times, 
the question arises about the proper ordering of such operators in the third
and higher order correlation functions. One can verify that the definition 
(\ref{cum}) corresponds to the following combination of the correlation 
functions:
\begin{eqnarray}
{\cal S}_3(t_1,t_2,t_2)&=&\frac{1}{8}
\big\{\big\langle\hat I(t_1)\big({\cal T}\hat I(t_2)\hat I(t_3)\big)\big\rangle
+\big\langle\big(\tilde{\cal T}\hat I(t_2)\hat I(t_3)\big)\hat I(t_1)\big\rangle
+\big\langle\hat I(t_2)\big({\cal T}\hat I(t_1)\hat I(t_3)\big)\big\rangle
+\big\langle\big(\tilde{\cal T}\hat I(t_1)\hat I(t_3)\big)\hat I(t_2)\big\rangle
\nonumber\\ &&
+\big\langle\hat I(t_3)\big({\cal T}\hat I(t_1)\hat I(t_2)\big)\big\rangle
+\big\langle\big(\tilde{\cal T}\hat I(t_1)\hat I(t_2)\big)\hat I(t_3)\big\rangle
+\big\langle{\cal T}\hat I(t_1)\hat I(t_2)\hat I(t_3)\big\rangle
+\big\langle\tilde{\cal T}\hat I(t_1)\hat I(t_2)\hat I(t_3)\big\rangle\big\}
\nonumber\\ &&
-\,\frac{1}{2}\big\langle\hat I(t_1)\big\rangle\big\langle\hat I(t_2)\hat I(t_3)+\hat I(t_3)\hat I(t_2) \big\rangle
-\frac{1}{2}\big\langle\hat I(t_2)\big\rangle\big\langle\hat I(t_1)\hat I(t_3)+\hat I(t_3)\hat I(t_1) \big\rangle
\nonumber\\ &&
-\,\frac{1}{2}\big\langle\hat I(t_3)\big\rangle\big\langle\hat I(t_1)\hat I(t_2)+\hat I(t_2)\hat I(t_1) \big\rangle
+2\big\langle\hat I(t_1)\big\rangle
\big\langle\hat I(t_2)\big\rangle\big\langle\hat I(t_3)\big\rangle.
\label{cum1}
\end{eqnarray}
This combination is of interest in the light of possible
experimental investigations of current fluctuations.
Indeed, the correlation function
${\cal S}_2$ is important because the symmetric combination
of voltages $V^+$ can be viewed as a classical, measurable,
voltage \cite{KN2}. The noise is deduced from the measurable
product $V^+(t_1)V^+(t_2),$ which is related to the symmetric
correlator ${\cal S}_2$ (\ref{corr}) or (\ref{noise1}). 
Analogously, the measured product
$V^+(t_1)V^+(t_2)V^+(t_3)$ is related to the correlation function
of the current operators ${\cal S}_3$ defined in Eq.
(\ref{cum1}), see also Refs. 14 and 20 for further discussion of this point.

Let us substitute the above expressions into Eq. (\ref{S3}) and,
after taking derivatives over $\varphi^-$, set $\varphi^+(\tau )=eV\tau$ and
$\varphi^- \to 0$. This is sufficient provided the time differences
$|t_1-t_2|,$ $|t_1-t_3|$ exceed the charge relaxation time $\tau_{RC}$ and
provided $eV \ll 1/\tau_{RC}.$ Eq. (\ref{S3}) then yields
\begin{eqnarray}
{\cal S}_3&=&\beta e^2\bar I\delta(t_1-t_2)\delta(t_1-t_3)-\frac{4\pi e\gamma
}{R} f(t_2-t_1) f(t_3-t_2)f(t_1-t_3),
\end{eqnarray}
where $f(\tau )=T\sin (eV\tau /2)/\sinh(\pi T\tau )$.
Performing the Fourier transformation
\begin{eqnarray}
{\cal S}_3(\omega_1,\omega_2)=\int d\tau_1 d\tau_2\,
{\rm e}^{i\omega_1\tau_1+ i\omega_2\tau_2}
{\cal S}_3(t_1,t_1-\tau_1,t_1-\tau_2),
\nonumber
\end{eqnarray}
we arrive at the final result
\begin{eqnarray}
{\cal S}_3=\beta e^2 \bar I -2\gamma e^2 \bar I F(v,w_1,w_2),\label{eq1}\\
F=\frac{\sinh^3(v/2)}{4v}\int_{-\infty}^\infty
 \frac{d\omega}{\chi (\omega )\chi (\omega-w_1) \chi (\omega+w_2)}.\label{funf}
\end{eqnarray}
Here we defined $v=eV/T,\,w_{1,2}=\omega_{1,2}/2T$ and
\begin{equation}
\chi (\omega )=\cosh^2\omega+\sinh^2(v/4).
\label{chi}
\end{equation}
Eqs. (\ref{eq1})-(\ref{chi}) represent the main result of this
section. They fully describe the third cumulant of the current
operator at voltages and frequencies smaller than both
$1/\tau_{RC}$ and $1/\tau_D$.

Let us briefly analyze Eqs. (\ref{eq1})-(\ref{chi}) in various limits. For
$\omega_{1,2}=0$ we recover the well known result \cite{lev2}
\begin{equation}
F(v,0,0)=1+3\frac{1-(\sinh v/v)}{\cosh v-1},\label{levv}
\end{equation}
which in turn yields $F\to 1$ in the limit of large voltages $v \gg 1$.
Eq. (\ref{levv}) also holds for $w_{1,2}\ll v$.

In the limit $v\ll 1$ one finds
\begin{eqnarray}
&&F(v\ll 1,w,0)=F(v\ll 1,w,-w)=\frac{9\sinh w+\sinh (3w)-12w\cosh
w}{48\sinh^5 w}v^2, \label{as1}
\end{eqnarray}
\begin{eqnarray} &&F(v\ll 1,w,w)=\frac{\sinh
(4w)+4\sinh(2w)-8w\cosh(2w)-4w}{128\sinh^5 w\cosh^3 w }v^2 .
\label{as2}
\end{eqnarray}
These equations demonstrate that at large frequencies $w \gg 1$  the function
$F$ decays exponentially with $w$. From Eqs. (\ref{as1}) and (\ref{as2})
we find respectively $F \propto v^2 e^{-2w}/3$ and $F \propto v^2 e^{-4w}$.

Finally let us turn to the most interesting limit of low temperatures, in which
case one always has $v,w_{1},w_2\gg 1$. Neglecting small corrections
$\sim 1/v,1/w$ we obtain
\begin{eqnarray}
F(v,w_1,w_2)= 1-2\left|\frac{w_{12}}{v}\right|,&&{\rm if }\;\;\;\;
2|w_{12}|<|v|,
\label{lt1}\\
F(v,w_1,w_2)= 0,&&{\rm if}\;\;\;\;2|w_{12}|>|v|. \label{lt2}
\end{eqnarray}
Here the value $w_{12}$ is defined differently depending on the sign of
the product
$w_1w_2$. For $w_1w_2>0$ we have $w_{12}=w_1+w_2$ while in the opposite case
$w_1w_2<0$ we define $w_{12}={\rm max}\,[|w_1|,|w_2|]$. We observe
that in both cases the function $F$ depends linearly on frequency and vanishes
as soon as $|w_{12}|$ exceeds $|v|/2$.

At arbitrary values of $v$, $w_1$ and $w_2$ the integral
(\ref{funf}) can be evaluated numerically. The corresponding
result for the function $F(v,w_1,w_2)$ is presented in Fig. 2.

\begin{figure}
\begin{center}
\includegraphics[width=7cm,angle=270]{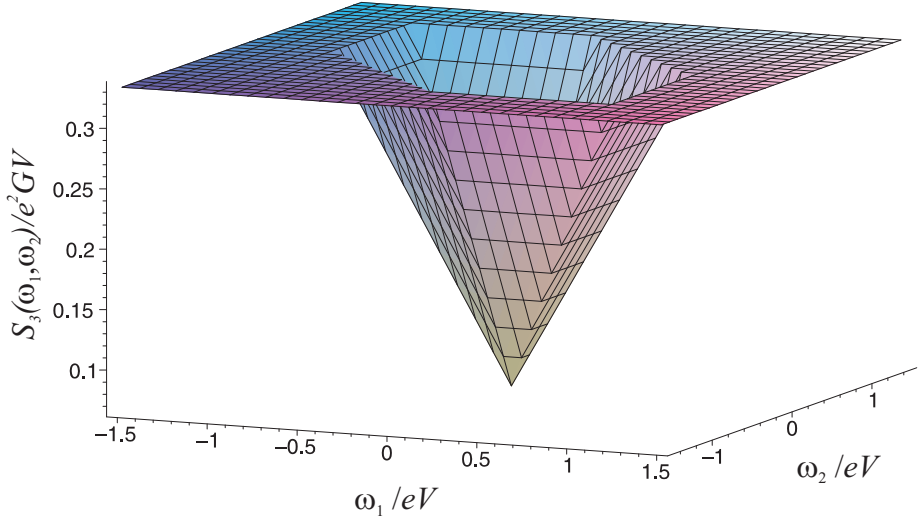}
\end{center}
\caption{Frequency dependence for the third cumulant of the current operator
${\cal S}_3(\omega_1,\omega_2)$ evaluated by means of Eqs. 
(\ref{eq1})-(\ref{chi}) at $T=0$ for a short diffusive 
conductor with $\beta=1/3$ and $\gamma=2/15$.}
\end{figure}

Let us emphasize again that in the course of
our analysis of the correlator ${\cal S}_3(\omega_1,\omega_2)$ we have ignored
electron-electron interactions and also assumed that the external impedance
$R_S$ equals to zero, i.e. the voltage source is directly attached to the
scatterer. In realistic experimental setups the latter condition may be violated, in which case the third cumulant can be substantially modified
\cite{been} by current fluctuations in the external environment (leads). 
Within our formalism this effect can be accounted for by the external impedance 
$Z_S(\omega)$.

\section{Current noise in the presence of interactions}

Let us now turn to the second cumulant (\ref{corr}) and explicitly
account both for interactions and the external leads $R_S$.
It is clear that fluctuations both in the scatterer and in the external
resistor will affect the correlator (\ref{corr}), i.e. the latter should in
general depend on both $R$ and $R_S$.
Let us -- in addition to (\ref{corr}) -- introduce another correlator $\tilde
{\cal S}(t,t')$
defined by the same Eq. (\ref{corr}) in which one should substitute
the current operator across the scatterer $\hat I \to \hat I_{\rm
  sc}$. The two
correlators  ${\cal S}(t,t')$ and $\tilde {\cal S}(t,t')$ are not independent.
With the aid of the current conservation condition and performing
the Fourier transformation with respect to $t-t'$, one easily obtains
the relation between these correlation functions:
\begin{eqnarray}
\tilde {\cal S}_\omega &=&\frac{R_S^2}{R_0^2}
(1+\omega^2R_0^2C^2){\cal S}_\omega
-\,\frac{R_S}{R^2}(1+\omega^2R^2C^2)\omega
\coth\frac{\omega}{2T},
\label{rel1}
\end{eqnarray}
where $R_0=RR_S/(R+R_S).$ The second term  is due
the noise produced by the external resistor $R_S$ which has to be
subtracted in order to arrive at $\tilde {\cal S}_\omega$.

In general also the correlator $\tilde {\cal S}_\omega$ depends on both $R$
and $R_S$. However, in the limit $R_S \gg R$ the dependence on the shunt
resistance is weak and can be neglected. In this case the interaction
correction to the current noise spectrum
depends only on the properties of the scatterer. Below we will present our
results only in this limit. More general
expressions can be found in Ref. 9.

As before, we will assume that the condition $g+g_S \gg 1$ is fulfilled
throughout our calculation. Combining the result (\ref{S3}) with the
definition (\ref{noise1}) and having in mind the relation (\ref{rel1})
one can directly evaluate the correlator  $\delta \tilde {\cal S}_\omega$
in the presence of interactions. In the important limit
$\omega,eV,T\ll 1/RC$ one finds
\begin{eqnarray}
S_2(\omega)&=& \frac{2(1-\beta)}{R}\omega\coth\frac{\omega}{2T}+ \beta e\tilde I\left(\frac{\omega}{e}+V\right)\coth\frac{\omega+eV}{2T}
+\beta e\tilde I\left(\frac{\omega}{e}-V\right)\coth\frac{\omega-eV}{2T}
\nonumber\\ &&
+\,\pi\gamma T^3 e^2\int_0^\infty dx\int_0^\infty dy\frac{(1-{\rm e}^{-x/RC})(\cos eVy-\cos eVx)\cos\omega y}{\sinh\pi Tx\sinh\pi Ty}
\nonumber\\ &&\times\,
\left(\frac{1}{\sinh\pi T(x-y)}-\frac{1}{\sinh\pi T(x+y)}\right),
\label{noise2}
\end{eqnarray}
where
\begin{equation}
\tilde I(V)=\frac{V}{R}-\frac{e}{\pi}\int_0^\infty dt\,\frac{\pi^2T^2}{\sinh^2\pi Tt}(1-{\rm e}^{-t/RC})\sin eVt.
\label{I}
\end{equation}
The result (\ref{noise2}) covers several important regimes. 
In the limit of weak tunneling $\beta\to 1$ and  $\gamma \to 0$ we find
\begin{eqnarray}
S_2(\omega)&=&  e\tilde I\left(\frac{\omega}{e}+V\right)\coth\frac{\omega+eV}{2T}
+ e\tilde I\left(\frac{\omega}{e}-V\right)\coth\frac{\omega-eV}{2T}
\label{noise3}
\end{eqnarray}
Here the function $I(V)$ is just the $I-V$ curve of a 
tunnel barrier modified by weak Coulomb blockade corrections. 
Eq. (\ref{noise3}) can be viewed as a generalization of the
result \cite{Dahm}. This formula also 
agrees with the result \cite{Lee} derived in the limit $g\ll 1.$

It is worth pointing out that Eq. (\ref{noise2}) can be reduced to the 
form (\ref{noise3}) only in the case of tunnel barriers. 
Otherwise the interaction correction to the current noise cannot
be obtained from the corresponding correction to the $I-V$ curve
except for equilibrium situations where the fluctuation-dissipation
theorem (FDT) can be employed.  
In order to illustrate this point and to present our results beyond the
tunneling limit let us split  the function $\tilde{\cal S}$ into two parts:
\begin{equation}
\tilde{\cal S}(t,t')=\tilde{\cal S}^{\rm ni}(t,t')+\delta \tilde{\cal S}(t,t'),
\label{int}
\end{equation}
where $\tilde{\cal S}^{\rm ni}$ is the noninteracting contribution to
the current noise
\cite{bb} and $\delta \tilde{\cal S}$ is the correction due to
electron-electron interactions inside the scatterer. 
Let us define the average voltage across the scatterer
$V=V_xR/(R+R_S)$ and consider first the limit of relatively small voltages.
At sufficiently large temperatures and/or frequencies we find
\begin{eqnarray}
\delta \tilde {\cal S}_\omega=-\frac{2\beta E_C}{3R},&&{\rm if}\;
 T\gg gE_C,|eV|,|\omega| ,
\label{highT} \\
\delta \tilde {\cal S}_\omega=-\frac{\beta E_C }{R},&&{\rm if}\;
|\omega|\gg T,gE_C,|eV|.
\end{eqnarray}

At lower temperatures and frequencies we obtain
\begin{eqnarray}
\delta \tilde {\cal S}_\omega=-\frac{4\beta T}{R_q}\ln\frac{gE_C}{T},
\;\;\;\;\;\;\;\;{\rm if}\; |\omega|,|eV|\ll T\ll gE_C,\label{intT}
\\ \delta \tilde {\cal S}_\omega=-\frac{2\beta |\omega|}{R_q}\ln
\frac{gE_C}{|\omega|}, \quad {\rm if}\; T,|eV|\ll |\omega|\ll gE_C .
\end{eqnarray}
These results apply as long as either temperature or frequency
exceeds the parameter $gE_C\exp (-g/2 )$.

As we have already pointed out, the above expressions could also be 
anticipated from FDT combined with the results
\cite{GZ00}. Indeed, in the limit of low voltages the current
noise is described by the standard Nyquist formula. Hence, in
order to satisfy FDT one should simply substitute the effective
conductance \cite{GZ00} into this formula. In this way one 
immediately reproduces Eqs. (\ref{highT}) and (\ref{intT}).

\begin{figure}
\begin{center}
\includegraphics[width=9cm]{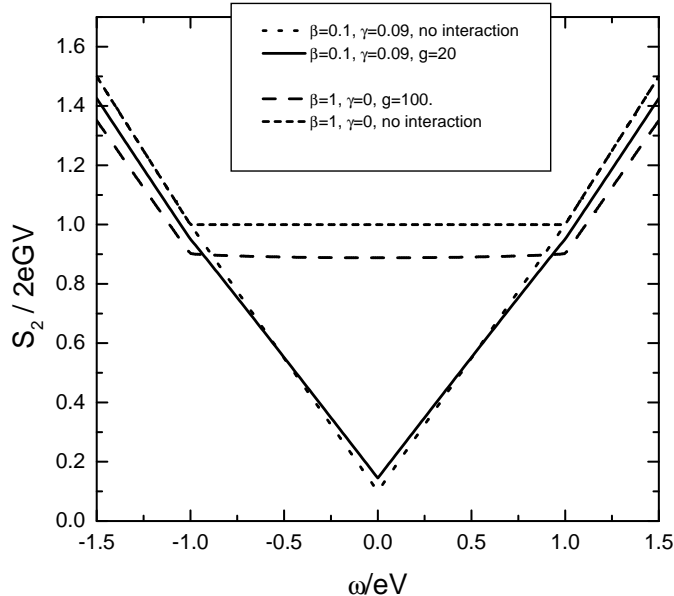}
\end{center}
\caption{Frequency dependence of the noise correlator ${\cal S}_2$ at $T=0$
and for various parameters of the system.}
\end{figure}

Now let us turn to the case of relatively large voltages $V$ where the
shot noise becomes important. In this case the correction to the noise 
power spectrum is found to be proportional to the parameter
\begin{equation}
\beta-2\gamma =\frac{\sum_nT_n(1-T_n)(1-2T_n)}{\sum_nT_n}.
\label{bg}
\end{equation}
We obtain
\begin{eqnarray}
\delta \tilde {\cal S}_\omega=-\frac{2(\beta-2\gamma)|eV|}{R_q}\ln
\frac{gE_C}{|eV|},\quad  {\rm if}\; T,|\omega|\ll |eV|\ll gE_C, \nonumber\\
\delta \tilde S_\omega=-\frac{(\beta-2\gamma)E_C}{R},\quad {\rm if}\;|eV|\gg
T,gE_C,|\omega|.
\end{eqnarray}
We note that this correction can be either negative or positive
depending on the relation between the parameters $\beta$ and $\gamma$.
Thus, in contrast to the limit of low voltages (Nyquist noise),
one {\it cannot} conclude
that shot noise is always reduced by interactions. This reduction
occurs only for conductors with relatively low transmissions $\beta >
2\gamma$, while for systems with higher transmissions the net effect
of the electron-electron interaction enhances the shot noise.
In the important case of diffusive conductors one has
$\beta =1/3$, $\gamma =2/15$ and, hence, $\beta -2\gamma =1/15$. In 
this case the shot noise is reduced by interactions.

The above results have a transparent physical interpretation. At low
voltages the power spectrum of the Nyquist noise is proportional to
the system conductance $\propto \sum_nT_n$. Since in the presence
of interactions the conductance acquires a correction proportional to
$\beta$, the interaction correction to the Nyquist noise should
scale with the same parameter. On the other hand, shot
noise is determined by the combination
$\sum_n(T_n-T_n^2)$. Accordingly, the interaction correction to the
shot noise power should consist of two contributions. One of
them comes from $\sum_nT_n$ and is again proportional to $\beta$.
Another contribution originates from the interaction correction
to  $\sum_nT_n^2$ which turns out to scale as $2\gamma$.
Since these two corrections enter with opposite signs we
immediately arrive at the combination (\ref{bg}).

We also point out that the third cumulant of the current operator for
noninteracting electrons is known \cite{lev2} to be proportional to the
parameters $\beta$ and $\beta -2\gamma$ respectively at low and high voltages.
Following the same arguments as above one can anticipate that the
interaction correction to the third cumulant should scale as $\beta -2\gamma$
at low voltages and as $\beta -6\gamma +6\delta$ at high voltages, where
$\delta =\sum_nT_n^3(1-T_n)/\sum_nT_n$
Technically, in order to determine the interaction correction to the third
cumulant ${\cal S}_3$ it is necessary to expand the exact effective action
$S_0$ (\ref{S2}) up to the fourth order in $\varphi^-$. Along with this
interaction correction such expansion
allows to determine the full frequency dependence of the fourth
cumulant ${\cal S}_4$ in the absence of interactions. 
This rule also applies to higher cumulants, i.e. the lowest order
interaction correction to the $n$-th current cumulant ${\cal S}_{n}$ is
determined by ${\cal S}_{n+1}$ for all values of $n$. This fact can also
be proven by means of the renormalization group analysis recently developed
in Refs. 11 and 12.


\section{Concluding remarks}

In this paper we have presented a general path integral approach which allows 
to describe
statistics of current fluctuations in mesoscopic coherent conductors at
arbitrary frequencies and in the presence of interactions. 
This approach enables one to establish a complete expression for
the effective action of coherent conductors described by an arbitrary --
though energy independent -- scattering matrix and to elucidate a 
profound relation between full counting statistics and 
electron-electron interaction effects in coherent mesoscopic conductors.
Further extention of our technique to more complicated structures described
by the energy dependent scattering matrices is possible and 
was recently worked out \cite{GZ03}. 

Restricting
ourselves to the non-interacting case, we have analyzed frequency dispersion
of the third cumulant of the current operator. This dispersion was found
negligible only in the case of tunnel junctions, while in a general case it
turns out to be important in the frequency range comparable to $eV$. 
For instance, in the important case of diffusive conductors and at $T=0$ 
the quantity
${\cal S}_3$ changes by the factor 5 depending on whether relevant
frequencies are below or above $eV$. For conductors with $\beta <
2\gamma$ even the sign of ${\cal S}_3$ differs in these two limits.

We have also analyzed the behavior of the second cumulant ${\cal S}_2$
in the presence of electron-electron interactions.
We have demonstrated that Coulomb interaction decreases the
Nyquist noise as one could anticipate already from the
results \cite{GZ00} combined with FDT. In the case of the shot noise, 
the effect of electron-electron interactions turns out to be more
complicated. The corresponding interaction correction was found
negative for conductors with $\beta > 2\gamma$ and positive otherwise.

Finally, let us point out that our predictions can be experimentally
tested in various types of coherent mesoscopic conductors, such as, e.g.,
break junctions, quantum point contacts or short
diffusive metallic bridges \cite{Weber,Sch}. In all these systems both
$\tau_D$ and $\tau_{RC}$ can be small enough in order to satisfy
all the assumptions adopted here. For instance, in diffusive samples
\cite{Weber,Sch} one finds $1/\tau_D$ of order few Kelvins and, hence,
the condition $eV < 1/\tau_D$ is obeyed in a wide range of voltages
$V < 0.1\div 0.5$ mV. We also note that the work \cite{Sch}
reports the experimental analysis of the frequency dispersion of
shot noise \cite{bb} which is important in the same frequency range as
that of the third cumulant studied here.

This work is part of the
Kompetenznetz ``Funktionelle Nanostructuren'' supported by the Landestiftung
Baden-W\"urttemberg gGmbH and of the STReP ``Ultra-1D'' supported by
the EU.

\end{document}